      [1999/12/01 v1.4c Il Nuovo Cimento]
\documentclass{cimento}

\usepackage[dvips]{graphicx}
\usepackage{epsfig,amsfonts}

\title{Neutrino Physics and $A_4$ matter assignments}
\author{
S.~Morisi\from{ins:x}}
\instlist{\inst{ins:x} Instituto de Fisica Corpuscular (IFIC) -
Centro Mixto CSIC-UVEG - Valencia, Spain}
\PACSes{%\PACSit
{11.30.Hv 14.60.-z 14.60.Pq 14.80.Cp}
}

\begin{document}

\maketitle

\begin{abstract}
Tribimaximal lepton mixing can be obtained assuming $A_4$ flavour symmetry.
Many possibilities have been explored in the literature and we give a classification in terms of $A_4$
representations. We propose some  phenomenological and theoretical criteria to distinguish between different assignment.
As example we consider the possibility to extend $A_4$ to $SO(10)$ grand unified model.
\end{abstract}

\section{Introduction}
It is an experimental fact that atmospheric, solar, reactor and accelerator
neutrinos change flavour. 
The present data \cite{Maltoni:2004ei} give the solar lepton mixing angle
$\sin^2 \theta_{12}=0.32$ with a relative error about of $25\%(@3\sigma)$, the atmospheric angle $\sin^2 \theta_{23}=0.5$ with a relative error 
about of $34\%(@3\sigma)$ and $\sin^2 \theta_{13}\le 0.05(@3\sigma)$. The complex phase has not been measured.
%Future reactor, Superbeam and SNOplus experiments will reduce the solar error only marginally, while more 
Precise future data might confirm the maximality of the atmospheric angle within $10\%$ error.
The best fit values agree very well with the so called Tri-Bimaximal (TB) 
lepton mixing matrix \cite{Harrison:2002er} where the atmospheric angle is maximal $\sin^2 \theta_{23}=1/2$, 
the solar angle is large  $\sin^2 \theta_{12}=1/3$ and  $\sin^2 \theta_{13}=0$. 
Assuming the charged lepton mass matrix $M_l$ diagonal, $U_{\mbox{\small TB}}\cdot {\mbox{Diag}}\left[m_1,m_2,m_3\right]\cdot U_{\mbox{\small TB}}^T$ takes
the form
\begin{eqnarray*}
M_\nu
%&=&U_{TB}\cdot Diag\left[m_1,m_2,m_3\right]\cdot U_{TB}^T=\\
&&=\left(
{%\small
\begin{array}{ccc}
\frac{2m_1}{3}+\frac{m_2}{3}& \frac{-m_1}{3}+\frac{m_2}{3}& \frac{-m_1}{3}+\frac{m_2}{3}\\
\frac{-m_1}{3}+\frac{m_2}{3}&\frac{m_1}{6} + \frac{m_2}{3} + \frac{m_3}{2}  & \frac{m_1}{6} + \frac{m_2}{3} - \frac{m_3}{2}\\
\frac{-m_1}{3}+\frac{m_2}{3}& \frac{m_1}{6} + \frac{m_2}{3} - \frac{m_3}{2} &\frac{m_1}{6} + \frac{m_2}{3} + \frac{m_3}{2}
\end{array}}
\right) \equiv
\left({\begin{array}{ccc} 
a&d&d\\
d&b&c\\
d&c&b
\end{array}}\right)
\end{eqnarray*}
Here $M_\nu$ has two important properties: {\it i)} it is $\nu_\mu\leftrightarrow \nu_\tau$ invariant, so $\theta_{13}=0$ and the atmospheric angle 
$\theta_{23}$ is maximal and  {\it ii)}  $a=b+c-d$ so from the relation
$$
\sin^2 2 \theta_{12}=\frac{8 b^2}{(a-b-c)^2+8 b^2}
$$
(that  is always true for  $\nu_\mu\leftrightarrow \nu_\tau$ invariant mass matrices \cite{Caravaglios:2005gw}) we have that $\sin^2{\theta_{12}}=1/3$.
The $\nu_\mu\leftrightarrow \nu_\tau$ invariance in the diagonal charged lepton basis {\it i)} can be explained with the 
permutation symmetry $S_3$ \cite{Grimus:2005mu},
while  the relation {\it ii)} is natural in $A_4$ models. 
\section{$A_4$ models and the selecting criteria}
$A_4$ is a finite group of the
even permutations of four objects.
%and it is isomorphic to the group of the tetrahedron symmetries. 
%$A_4$ has four irreducible representation, three distinct singlets $1,1',1''$  and a triplet $3$ and
$A_4$ is the smallest non-Abelian group that contains a triplet representation. 
It also contains three distinct singlets $1,1',1''$.
%This is a nice feature since 
We can accommodate
the three families of fermions both in a triplet and/or in the three singlets representations. In Table \ref{tab1} we report
all possible assignment with at least one triplet representation.  
%n literature has been studied models where matter is differently assigned into
%rreducible representations of $A_4$ (see Table \ref{tab1}). 
For instance, in ref.\,\cite{Babu:2002dz} $A_4$ leads to a $\nu_\mu\leftrightarrow \nu_\tau$ invariant neutrino mass matrix, 
therefore $\theta_{13}=0$, $\theta_{23}$ is maximal and $\theta_{12}$ can be fitted within the experimental error. 
Recently in \cite{Altarelli:2005yp} it has been  studied a model that predict also the solar angle.
To distinguish these models we need some selecting criteria.
\begin{table}[t]
\begin{tabular}{lrcccrcc}
\hline
SM& & $L_i $ & $l_{i}^c$ &$\nu^c_i$ &References\\
\hline
type-I&I  &3&$1,1',1''$&3 & \cite{Babu:2002dz,Altarelli:2005yp,Ma:2002yp,1}\\
%&II  &3&$1,1',1''$& -&\cite{2}\\
%&III  &3&3& -& \cite{3}\\
&II &3&3&$1,1',1''$ &\cite{4}\\
&III &$1,1',1''$&3&3&\cite{5}\\
&IV &$1,1',1''$&$1,1',1''$ &3&-\\
&V &$1,1',1''$&3&$1,1',1''$ &-\\
&VI &3&$1,1',1''$&$1,1',1''$ &-\\
\hline
type-II&I  &3&$1,1',1''$& -&\cite{2}\\
&II  &3&3& -& \cite{3}\\
&III  &$1,1',1''$ &3& -& -\\
\hline
\end{tabular}\caption{Different $A_4$ matter assignments for type-I and II seesaw.}\label{tab1}
\end{table}
One possibility could be to study their {\it phenomenological} implication:

- the  neutrinoless double beta decay rate and the leptonic CP phase;

%- deviations from TB mixing;

- the stability under radiative corrections and the deviations from TB mixing;

%- predictively (number of free parameters); 

- LHC phenomenology, for instance, %some of the $A_4$ models introduce more than one 
Higgs doublets and/or Higgs triplets. \\
From the {\it theoretical} point of view we have at least two general criteria: 

- extend $A_4$ to the strong sector without spoiling the TB mixing: explicitly breaking 

\quad $A_4$ \cite{Ma:2002yp}; assign $q_L$ and $q_R$ 
   differently to $A_4\times Z_{n_1}\times...Z_{n_k}$ \cite{volkas}; extra dimensions \cite{Altarelli:2008bg};

- extend the $A_4$ symmetry to grand unified models (GUT) \cite{Altarelli:2008bg,King:2006np,Morisi:2007ft}. 
\section{$A_4$ and $SO(10)$ grand unified model}
\begin{figure}[t]
\begin{center}
\includegraphics[angle=0,width=60mm]{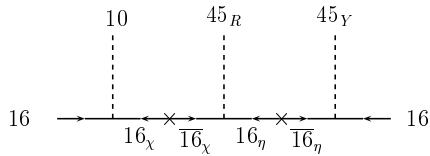}\caption{Six dimension operator distinguishes up quarks and Dirac neutrinos.}\label{fig1}
\end{center}
\end{figure}

%Hints toward the existence of GUT are the possible unification of the SM gauge couplings,
%the explanation of the charge quantization and the anomaly cancellation.
In  $SO(10)$ all the SM matter fields belong to one 16-multiplet. Neutrinos
get small masses in a natural way through the seesaw mechanism since the right-handed neutrinos
get  Majorana masses at the unification scale while the Dirac masses can have values at the electroweak
scale. $SO(10)$ forces to assign left and right handed fields in one triplet of $A_4$ giving strong constraints in the  model building.
First we give an example at the electroweak scale compatible with such a matter assignment. Consider the following
\begin{center}
\begin{small}
\begin{tabular}{cccccccc}
\hline
SM & $L_{i} $ & $l^c_i$ &$N^c_i$  & $\phi$&$\phi'$&$\xi_1$&$\xi_2$\\
\hline
$A_4$ & 3& 3 & 3 &3&3&1&1 \\
$Z_3$ &$\omega$&$\omega$ &$\omega^2$ &$\omega$&$\omega^2$&$\omega$&$\omega^2$\\
\hline
\end{tabular}
\end{small}
\end{center}
At the leading order $\phi$ and $\phi'$ interact respectively only with the charged and neutral sectors.
When $\langle\phi\rangle\sim (1,1,1)$, $M_l$ is diagonalized from the magic matrix $U_\omega$, see \cite{4}.
If  $\langle\phi'\rangle\sim (1,0,0)$, in the diagonal charged lepton basis we get
%\begin{equation}
$$
M_\nu=m_D\frac{1}{M_R}m_D^T\equiv
\begin{small}
\left({\begin{array}{ccc} 
2b/3  +a& - b/3 &-b/3\\
- b/3  &2b/3  &- b/3 +a\\
- b/3 &- b/3 +a& 2b/3 
\end{array}}\right)
\end{small}
$$
%\end{equation}
that has the properties {\it i)} and {\it ii)}. Here we have used the relation $m_D\sim I$ that is a consequence of the 
model. 
Such a relation could be a problem since in $SO(10)$ we expect $m_D\sim M_u\sim I$ that is wrong. This argument seems against 
$SO(10)\times A_4$. In \cite{Morisi:2007ft} we have studied the possibility to distinguish up quark and Dirac neutrino
through dimension six operators showed in Fig.(\ref{fig1}) where $16_{\chi,\eta}$s are extra messenger fields. When the adjoints $45_R$ and $45_Y$ take vev respectively along weak isospin and 
the hypercharge, such operator do not contribute to the Dirac neutrino.

\acknowledgments

Work supported by MEC grant FPA2005-01269 and FPA2005-25348-E, by Generalitat Valenciana ACOMP06/154, by European Commission Contracts MRTN-CT-2004-
503369 .
%by European Commission Contracts MRTN-CT-2004-
%503369 and ILIAS/N6 RII3-CT-2004-506222.

%The organizers......

\end{document}